\documentstyle[12pt]{article}
\def\1{\mbox{I\hspace{-.15em}1}}

\def\b{\begin{equation}}
\def\e{\end{equation}}
\def\bee{\begin{enumerate}}
\def\eee{\end{enumerate}}

\title{Friedmann-like equations
\centerline{For}
\centerline{High Energy Area of Universe}}
\author{E. Yusofi$^{1}$\thanks{e-mail:
e.yusofi@iauamol.ac.ir} and M. Mohsenzadeh$^{2}$\thanks{e-mail:
mohsenzadeh@qom-iau.ac.ir}}
\begin{document}

\maketitle {\it   \centerline{$^1$
Department of Physics, Ayatollah Amoli Branch, Islamic Azad University, Iran}
\centerline{\it P.O.BOX 678, Amol, Mazandaran} \centerline{\it $^{2}$ Department of Physics, Qom
Branch, Islamic Azad University, Qom, Iran }}

\begin{abstract}In this paper, evolution of the high energy area of universe, through the scenario of 5 dimensional (5D) universe, has been studied. For this purpose, we solve Einstein equations for 5D metric and 5D perfect fluid to derive Fridmann-like equations. Then we obtain the evolution of scale factor and energy density with respect to both space-like and time-like extra dimensions. We obtain the novel equations for the space-like extra dimension and show that the matter with zero pressure cannot exist in the bulk. Also, for dark energy fluid and vacuum fluid, we have both accelerated expansion and contraction in the bulk.

\end{abstract}
Keywords: Extra Dimensions, Friedmann Equations, High Energy Physics.
PACS numbers:  98.80.-k, 04.50.+h, 11.25.Mj


\section{Introduction and Motivation}
The most successful model, having wondrous predictive power for cosmology, was obtained by Friedmann in 1922 \cite{c1}, as well as by Robertson and Walker in 1935 and 1936, respectively \cite{c2,c3}, which is called as Friedmann-Robertson-Walker (FRW) model. This expanding and centrally symmetric homogeneous model obeys the cosmological principle. The successfulness of this model is revealed by the fact that the big-bang model is based on this model \cite{c4,c5}.

After Hubble's observation, in 1929, Einstein realized that he committed a mistake by introducing the cosmological constant. Due to this epoch-making observation, cosmologists were prompted to think for expanding models of the universe more seriously, though de-Sitter and Friedmann had already derived non-static models by this time. So, it was natural to think that if the present universe is so large, it would have been very small and dense in the extreme past. In 1940, George Gamow addressed to this question and proposed that, in the beginning of its evolution, universe was like an extremely hot ball \cite{c6}. He called this ball as primeval atom and this model is popularly known as the big-bang model.

The idea of more than three spatial dimensions has a rich history \cite{c7}. During the twentieth century it got remarkable recognition when Kaluza \cite{c8} and Klein \cite{c9} tried to use the extra spatial dimension for unifying the gravity with other forces. Later on, Witten \cite{c10} attempted to unify the weak and strong forces with gravity on a purely theoretical ground and generalized the findings to higher dimensions. The use of extra dimensions in describing the theories of the cosmology as well as those of particle physics has paved a path to find the solution of some of the fundamental questions in the micro and macro level physics. The whole visible universe now is supposed to be stuck on a membrane, in this higher-dimensional space-time, like a dust particle on the shop bubble. As such, a number of theoretical physicists and the cosmologists are developing the theoretical models of creation of matter and energy at micro and macro level using extra dimensions. Attempts are being made to explain different aspects of the events in the universe and to interpret the existence of energy and matter from atom to galaxy with more than four-dimensional space-times. One reason of this endeavor is to develop the model for the unification of all the four fundamental forces and then ultimately find a theory of everything.

Most of the earlier studies on the higher dimensions were based on the smallness or compactification of extra dimensions. Role of extra dimensions in the superstring theories, which are supposed to be most accepted theories of unification, was based on the idea of compactification. However, in the perspective of cosmology the role of extra dimension seems to be changed. In the present epoch most of the experimental results in cosmology reveal that the universe is not expanding with a uniform speed but it is showing the accelerated expansion \cite{c11,c12}. This has not only changed our earlier belief of a retarded expansion but also forced the cosmologists to modify the theories accordingly. Therefore, since the last decade much attention is being paid on explaining the cause of acceleration of the universe in the late time \cite{c13,c14,c15,c16}. However, we have no definite clue of the cause of acceleration.

Attempts have been made to explain the acceleration of the universe with the help of the cosmological constant (vacuum energy), dark energy associated with scalar field, spatial extra dimension, superstring, branes and modifying the general relativity at low energy \cite{c17,c18,c19,c20,c21}.

The experimental findings of the accelerated expansion of the universe, with higher-dimensional cosmological theories, have given a new thinking to the modern physicists. The extra dimension is supposed to play important role in this acceleration. Also, cosmology with homogeneous extra dimension has been found to describe by generalized FRW metrics to find the field equations in static extra dimensions, both in the case of radiation dominated and the cosmological constant dominated space-time \cite{c22}. Some exact solutions of the cosmological models in higher dimensions with the time dependent energy momentum has been investigated by Chatterjee \cite{c15}.

Gu et al \cite{c23} studied the universe with homogeneous extra dimension, and found an evolution pattern in the radiation dominated era and the accelerating expansion was found with a variable G. Chatterjee \cite{c15} also agreed on the suggestion that one of the major cause of the acceleration of the universe is the presence of the extra spatial dimension in the four-dimensional space-time of the physical theories, without any external factor. Therefore, he discussed a scenario in homogeneous 5D space-time which admits a decelerating expansion in the early epoch along with an accelerated phase at present. The case of obtaining acceleration and the inflation from higher-dimensional gravitational theory is undertaken by Ohta \cite{c24}. Purohit and Bhatty \cite{c25} considered a five-dimensional FRW-type Kaluza-Klein cosmological model to study the role of extra dimension in the expansion of the universe. Tiwari et al \cite{c26} took a five-dimensional Kaluza-Klein space-time in the presence of a perfect fluid source with variable G and which provides an expanding universe. Also we showed that the extra dimension can produce the negative pressure in the early five-dimensional universe \cite{c27}.

Because the space-time with higher dimensions may have more energy than the space-time with 4 dimensions, in this paper, we suggest that the primeval atom of the Gamow idea has additional dimension (dimensions) and we examine evolution of this 5D universe respect to additional dimension variables. Be careful that our scenario is not only as a model of cosmology with an extra dimension, but more generally, we use the methods of the standard cosmology as a tool to investigate the evolution of high energy regions of physics. So In Sect.2, the solution of Einstein equations with the standard Robertson-Walker (RW) metric is recalled briefly. In sect.3, we drive Fridmann-like equations for 5D universe. In sect.4, we obtain the evolution of the Bulk scale factor \cite{c28} and bulk energy density. Also, we consider the different shapes of perfect fluid to describe the physical significance of the new relations of our scenario. The results are discussed in final section.

\section{Freidmann Equations for 4D Universe }
In comoving coordinates, the cosmological space-time is given by
\cite{c4,c5}, \b
ds^{2}=dt^{2}-a^{2}(t)[dx_{1}^{2}+dx_{2}^{2}+dx_{3}^{2}],\e where
$ a(t) $  is called the scale factor.\\ At cosmic scale, the
contents of the universe look like non-interacting tiny gas
particles having homogeneous and isotropic distribution. So, the
cosmic matter behaves like a perfect fluid and its energy-momentum
tensor is given by, \b
T^{\mu}_{\nu}=\textit{diag}(\rho,-P,-P,-P),\e and the equation of
state for this fluid is, \b P=\omega \rho,\e where $ \rho $ and $
P $ are energy and pressure respectively.

On the other hand, Einstein's field equations are given by, \b
R_{\mu\nu}=-8\pi G (T_{\mu\nu}-\frac{1}{2}g_{\mu\nu}T).\e Thus,
for non-vanishing components of $ R_{\mu\nu} $ and $ T_{\mu\nu} $,
we have from (4), \b (\frac{\dot{a}}{a})^{2}=\frac{8 \pi
G}{3}\rho,\e where dot denotes derivative with respect to time.\\
The Bianchi identities are given by, \b \nabla_{\nu} T^{\mu\nu}=0,
\e which can be rewritten as, \b
\frac{1}{\sqrt{-g}}\frac{\partial}{\partial
x^{\nu}}(\sqrt{-g}T^{\mu\nu})+\Gamma^{\mu}_{\alpha \nu}T^{\alpha
\nu}=0. \e

In the RW space-time, for non-vanishing components of $
\Gamma^{\mu}_{\alpha \nu} $ and $ T^{\alpha \nu} $, we have,\b
\dot{\rho}+3H(\rho+P)=0. \e This is the conservation equation for
the homogeneous perfect fluid constituting the cosmic fluid. It is
just enough to choose the set of Eqs. (5) and (8) for our work
in cosmic evolution in FRW cosmology. Thus, from $ P=\omega \rho
$, the conservation of energy equation (8) becomes,\b
\frac{\dot{\rho}}{\rho}=-3(1+\omega)\frac{\dot{a}}{a}.\e This can
be integrated to obtain,\b \rho=a^{-3(1+\omega)}.\e We then
find from (5) and (10), the expansion of the universe goes as
\cite{c29}, \b a\approx t^{\frac{2}{3(1+\omega)}}. \e In
particular, for dark energy fluid with $ -1<\omega<-\frac{1}{3} $,
we obtain \b \rho\approx a^{m},\e and \b a\approx t^{n}.\e where $
-2<m<0 $ and $ 1<n<\infty $ . In this case, the energy density of
the universe decrease and  the acceleration of the universe
increase $ (a>0) $. Also for  $ \omega=-1 $  or vacuum, we
have \b \rho\approx a^{0}=cte.\e In this case the energy density
is independent of $ a $, but we have inflation for the scale
factor as, \b a\approx t^{\infty}.\e Thus, in any case, the scale
factor vanishes if $ t\longrightarrow 0 $ and the density at that
time becomes infinite. This point is known \textit{Big Bang}.

\section{ Freidmann-Like Equations for 5D Universe }
For our scenario, we consider the warped metric in braneworld scenario model \cite{c30} as the modified form of  RW metric in higher dimensional universe. So, the line element is given by
$$ ds^{2}=g_{ab}dx^{a}dx^{b}=a^{2}(y)\eta_{\mu\nu}dx^{\mu}dx^{\nu}+\epsilon^{2}dy^{2}.$$
\b g_{ab}=\textit{diag}(a^{2}(y),-a^{2}(y),-a^{2}(y),-a^{2}(y),\epsilon^{2}).\e
where $ a(y) $ is the warp factor and it is supposed to be a real function of the
extra dimension, which gives rise to the warped geometry and $ \eta_{\mu\nu} $ describes the 4D flat space-time, with $ \mu,\nu=0,1,2,3 $. We use $ \epsilon^{2}=-1 $ for the Space Like Extra Dimensions (SLED) and
$ \epsilon^{2}=+1 $ for the Time Like Extra Dimensions (TLED). The 5D metric given above, implies that the scalar of curvature and the warp factor only depend on the fifth coordinate y, i.e. $ R\equiv R(y) $, and $ a \equiv a(y) $ respectively.

We consider two different shapes of the 5D energy-momentum tensor for SLED and TLED separately. The energy-momentum tensor for SLED
reads as, \b T^{a}_{b}=
\textit{diag}(\rho,-P,-P,-P,-\tilde{p}).\e and for TLED is
\b T^{a}_{b}=
\textit{diag}(\rho,-P,-P,-P,\tilde{\rho}).\e
Because of the homogeneity condition, we consider $ \tilde{p}= P $ and $ \tilde{\rho}=\rho $, as the pressure and
density of the fifth component of prefect fluid energy-momentum tensor.

Similar to previous section, we solve Einstein's field equations in the 5D
space-time (16),\b R_{ab}=-8\pi G (T_{ab}-\frac{1}{2}g_{ab}T).\e
Thus, for non-vanishing components of $ R_{ab} $ and $ T_{ab} $,
we obtain from (19) for $ ab=44 $, in SLED case as, \b
\frac{\tilde{a}''}{\tilde{a}}=\pi G(2P-\rho). \e where $
\tilde{a}\equiv a(y) $ is bulk scale factor, and prime
denotes derivative with respect to $ y $ as a SLED. Also for $
ab=(11,22,33) $, we obtain from (19), \b
\frac{3\tilde{a}'^2}{\tilde{a}^2}+\frac{\tilde{a}''}{\tilde{a}}=-4\pi
G\rho. \e

With connecting Eqs.(20) and (21), we obtain,\b \frac{\tilde{a}'^2}{\tilde{a}^2}=\frac{8 \pi G}{3}(\frac{-3\rho}{8}-\frac{P}{4}). \e

On the other hand, similar to SLED case, for TLED we obtain,\b \frac{\ddot{\tilde{a}}}{\tilde{a}}=-3\pi G P, \e and
\b \frac{3\dot{\tilde{a}}^2}{\tilde{a}^2}+\frac{\ddot{\tilde{a}}}{\tilde{a}}=4\pi G (2\rho - P). \e where dot denotes
derivative with respect to $ y $ as a TLED. with connecting Eqs. (23) and (24),
we obtain,\b \frac{\dot{\tilde{a}}^2}{\tilde{a}^2}=\frac{8\pi G}{3} (\frac{P}{8} - \rho). \e
Also, from the Bianchi identities,$ \nabla_{b} T^{ab}=0 $, we obtain the conservation equation
for SLED as,\b P'+\frac{\tilde{a}'}{\tilde{a}}(\rho+P)=0, \e and for TLED,\b \dot{\rho}+3\frac{\dot{\tilde{a}}}{\tilde{a}}(\rho+P)=0. \e

\section{Evolution of Bulk Scale Factor and bulk energy density in 5D Universe}
We would like give the evolution of bulk scale factor $ \tilde{a} $ respect to variation of $ y $ in the bulk. Therefore, if we consider
 $ P=\omega \rho $, with integration of (26) for SLED, the conservation of energy equation becomes,
 \b \frac{\rho'}{\rho}=-\frac{(1+\omega)}{\omega}\frac{\tilde{a}'}{\tilde{a}},\e This can be integrated to obtain,
 \b \rho\approx\tilde{a}^{-(1+\frac{1}{\omega})}, \e We then find from (22) and (29), the evolution of the bulk scale factor goes as,\b \tilde{a}\approx y^{\frac{2\omega}{(1+\omega)}}, \e and for
 TLED case we obtain similar to 4D standard universe, i.e.,\b \rho\approx\tilde{a}^{-3(1+\omega)}, \e
 and \b \tilde{a}\approx y^{\frac{2}{3(1+\omega)}}. \e Although, the relations (31) and (32) are similar
 to (10) and (11), but for SLED case we obtain the new relations (29) and (30).

We shall now consider particular values of $ \omega $ to describe the physical
significance of the new relations (29) and (30). Therefore, we obtain for:

$ \bullet $	Matter dominate $ \omega=0 $

$$ \rho \propto a^{-\infty}\rightarrow 0 $$
\b a\propto y^{0}=const\e
In SLED case, the matter (the cold non-relativistic matter) cannot exist in the bulk. This result seems quite reasonable because the SLED is an ultra-relativistic region with a velocity faster than velocity of light. But for TLED, evolution of bulk scale factor and bulk energy density is similar to the standard cosmology.

$ \bullet $	Radiation dominate $ \omega=\frac{1}{3} $

$$ \rho \propto a^{-4} $$
\b a\propto y^{\frac{1}{2}}\e
For both SLED and TLED, evolution of scale factor and energy density for radiation (ultra- relativistic matter) in the bulk is similar to the evolution of them in the standard cosmology.

$ \bullet $	Vacuum dominate $ \omega=-1 $

$$ \rho \propto a^{0}=const $$
\b a\propto y^{-\infty}\rightarrow 0\e

$ \bullet $	Dark energy fluid $ -1<\omega<\frac{-1}{3} $

$$ \rho \propto a^{2>m>0}=const $$
\b a\propto y^{-1>n>-\infty}\e
In this case for vacuum and dark energy fluid, the bulk scale factor is similar as warp factor in braneworld scenario model and we have accelerated contraction in the bulk, but for TLED case, the bulk scale factor behave similar as scale factor in FRW model and we have accelerated expansion in the bulk.

\section{Discussions and Conclusions}
We considered the warped metric in braneworld scenario model as the modified form of RW metric in higher dimensional universe, which admits both SLED and TLED. We solved the Einstein equation for this space-time and demonstrated that the form of evolution of the bulk scale factor and bulk energy density for SLED case is new. Also for dark energy fluid and vacuum fluid we have both accelerated expansion and contraction in the bulk. This means two types of extra dimensions seems to be important in this scenario and play two different roles.\\
Moreover, the following results and points can be considered:

 1.
In this higher dimensional scenario, with incorporation of hidden extra dimensions and the visible space-time dimensions, the unification of gravitational force and gauge forces seems to be more possible.

2. In this scenario, for TLED we have an accelerated expansion of bulk scale factor and the strong gravitational force in the bulk is likely repulsive, but for SLED we have an accelerated contraction of the bulk scale factor (35) and the strong gravitational force in the bulk is likely attractive. Consequently, in the bulk, the strong gravitational force can be attractive as well as repulsive. Thus, the quantum effects of gravitation are of great importance in the bulk and in the higher dimensional universe \cite{c31}.

The accelerated contraction of bulk scale factor (35) can be similar to the process of shrinking of space-time in the formation of black holes, but the accelerated expansion of bulk scale factor can be similar to processes such as inflation in the early universe or exploding of space-time in supernovas. Consequently, for the former, evolution of bulk scale factor can be end with a "big destruction" and for the latter; the evolution can be started with a "big explosion". With acceptation these similarities, we conclude that extra dimensions may play a fundamental role in the high energy physics events.

\noindent {\bf{Acknowlegements}}:This work has been supported by the Islamic Azad University, Ayatollah Amoli Branch, Amol, Iran.


\begin{thebibliography}{a}

\bibitem{c1}Friedmann, A.Z.: Physik 10, 377 (1922).
\bibitem{c2}Robertson, H.P.: Astrophys. J. 82, 284 (1935).
\bibitem{c3}Walker, A.G.: Proc. Lond. Math. Soc. 42(2), 90 (1936).
\bibitem{c4}Carroll, S.M.: An Introduction to General Relativity: Spacetime and Geometry. Addison Wesley, Reading (2004).
\bibitem{c5}Srivastava, S.K.: General Relativity and Cosmology. PHI (2008).
\bibitem{c6}Gamow, G.: Nature 162, 680 (1948).
\bibitem{c7}H. P. Manning, The Fourth Dimension Simply Explained (Dover Publication, New York, 1910).
\bibitem{c8}T. Kaluza, Sitzungsber. Preuss. Akad. Wiss. Berlin (Math. Phys.) K 1, 966 (1921).
\bibitem{c9}O. Klien, Z. Phys. 37, 895 (1926).
\bibitem{c10}E. Witten, Phys. Lett. B 144, 351 (1984).
\bibitem{c11}A. G. Riess et al., Astron. J. 116, 1009 (1998).
\bibitem{c12}S. Perlmutter, Astrophys. J. 517, 565 (1999).
\bibitem{c13}R. R. Caldwell, R. Dave and P. J. Steinhardt, Phys. Rev. Lett. 80, 1582 (1998).
\bibitem{c14}J. A. Gu, arXiv:astro-ph/0209223.
\bibitem{c15}S. Chatterjee, A simple accelerating model; of the universe in higher dimensional
space-time, in 22nd Texas Symp. on Relativistic Astrophysics at Stanford University,
13-17 December 2004, eds. P. Chen et al. (Stanford University Publication, 2004).
\bibitem{c16}J. A. Gu and W. Y. P. Hwang, Phys. Rev. D 66, 024003 (2002).
\bibitem{c17}S. M. Carroll, V. Duvvuri, M. Trodden and M. S. Turner, Phys. Rev. D 70, 043528 (2004).
\bibitem{c18}D. A. Easson, Int. J. Mod. Phys. A 19, 5343 (2004).
\bibitem{c19}N. Arkani-Hamed, S. Dimopolos and G. Dvali, Phys. Lett. B 429, 263 (1998).
\bibitem{c20}N. Arkani-Hamed, S. Dimopolos and G. Dvali, Phys. Rev. D 59, 086004 (1999).
\bibitem{c21}L. Randall and R. Sundrum, Phys. Rev. Lett. 83, 3370 (1999).
\bibitem{c22}T. Bringmann and M. Ericksson, Cosmol. Astropart. Phys. J. 10, 1 (2003).
\bibitem{c23}J. A. Gu, W. Y. P. Hwang and J. W. Tsai, arXiv:astro-ph/0403641.
\bibitem{c24}N. Ohta, arXiv:hep-th/0411230.
\bibitem{c25}K. D. Purohit and Y. Bhatty, Int J Mod Phys A(2008) 909- 917.
\bibitem{c26} R.K. Tiwari , F.  Rahaman , S.  Ray, Int J Theor Phys (2010) 49: 2348-2357.
\bibitem{c27}Yusofi, E., Mohsenzadeh, M.:Int. J. Theor. Phys. 49, 1556-1561 (2010).
\bibitem{c28}Mohsenzadeh, M., Yusofi, E.:Int. J. Theor. Phys. 50, 430-435 (2011).
\bibitem{c29}A. Linde: Particles Physics and Inationary Cosmology. Harwood Academic, Reading
(1991); A.R. Liddle: An introduction to cosmological ination. arXiv:astro-ph/9901124v1 (1999).
\bibitem{c30}V.I. Afonso, D.Bazeia, R. Menezes, and A.Yu. Petrov, Phys. Lett. B 658, 71-76 (2007) [arXiv:hep-th/0710.3790].
\bibitem{c31}K. Nozari, , S.H. Mehdipour,: Int. J. Mod. Phys. A 21, 4979-4992 (2006).


\end{thebibliography}
\end{document}